\documentclass[aps,prl,twocolumn,superscriptaddress,reprint]{revtex4-2}
\usepackage{graphicx}
\usepackage{amssymb, amsmath, amsthm}
\usepackage{booktabs}
\usepackage[usenames]{color}
\usepackage{bm}
\usepackage{multirow}
\usepackage{dcolumn}
\usepackage{hyperref}
\usepackage{enumerate}
\usepackage{siunitx}
\usepackage[mathlines]{lineno}
\hypersetup{
    colorlinks=true,
    linkcolor=blue,
    filecolor=gray,      
    urlcolor=blue,
    citecolor=blue,
}

\begin{document}

\title{Improving density matrix electronic structure method by deep learning}

\newcommand{\thuphy}{State Key Laboratory of Low Dimensional Quantum Physics and Department of Physics, Tsinghua University, Beijing 100084, China}
\newcommand{\thuias}{Institute for Advanced Study, Tsinghua University, Beijing 100084, China}
\newcommand{\fscqi}{Frontier Science Center for Quantum Information, Beijing 100084, China}
\newcommand{\riken}{RIKEN Center for Emergent Matter Science (CEMS), Wako, Saitama 351-0198, Japan}
\newcommand{\pkuchem}{College of Chemistry and Molecular Engineering, Peking University, Beijing 100871, China}

\affiliation{\thuphy}
\affiliation{\thuias}
\affiliation{\fscqi}
\affiliation{\riken}
\affiliation{\pkuchem}

\author{Zechen \surname{Tang}}
\thanks{These authors contributed equally to this work.}
\affiliation{\thuphy}

\author{Nianlong \surname{Zou}}
\thanks{These authors contributed equally to this work.}
\affiliation{\thuphy}

\author{He \surname{Li}}
\affiliation{\thuphy}
\affiliation{\thuias}

\author{Yuxiang \surname{Wang}}
\affiliation{\thuphy}

\author{Zilong \surname{Yuan}}
\affiliation{\thuphy}

\author{Honggeng \surname{Tao}}
\affiliation{\thuphy}

\author{Yang \surname{Li}}
\affiliation{\thuphy}

\author{Zezhou \surname{Chen}}
\affiliation{\thuphy}

\author{Boheng \surname{Zhao}}
\affiliation{\thuphy}

\author{Minghui \surname{Sun}}
\affiliation{\thuphy}

\author{Hong \surname{Jiang}}
\affiliation{\pkuchem}

\author{Wenhui \surname{Duan}}
\email{duanw@tsinghua.edu.cn}
\affiliation{\thuphy}
\affiliation{\thuias}
\affiliation{\fscqi}

\author{Yong \surname{Xu}}
\email{yongxu@mail.tsinghua.edu.cn}
\affiliation{\thuphy}
\affiliation{\fscqi}
\affiliation{\riken}

\begin{abstract}
The combination of deep learning and ab initio materials calculations is emerging as a trending frontier of materials science research, with deep-learning density functional theory (DFT) electronic structure being particularly promising. In this work, we introduce a neural-network method for modeling the DFT density matrix, a fundamental yet previously unexplored quantity in deep-learning electronic structure. Utilizing an advanced neural network framework that leverages the nearsightedness and equivariance properties of the density matrix, the method demonstrates high accuracy and excellent generalizability in multiple example studies, as well as capability to precisely predict charge density and reproduce other electronic structure properties. Given the pivotal role of the density matrix in DFT as well as other computational methods, the current research introduces a novel approach to the deep-learning study of electronic structure properties, opening up new opportunities for deep-learning enhanced computational materials study.
\end{abstract}

\maketitle

\section{Introduction}

Deep-learning-enhanced ab initio calculations have emerged as powerful tools for accelerating materials property calculations, and have demonstrated efficacy in expediting materials discovery, facilitating molecular dynamics simulations, modeling electronic structure calculations, and computing various physical quantities~\cite{Behler2007,Zhang2018,Schutt2019,Unke2021,deeph2022,deeph-e32023,xdeeph2023,deeph-dfpt2024,deeph-hybrid2023,deeph22024,deephumm2024,magnet2024,yu2023efficient}. By circumventing the time-consuming self-consistent iterations in DFT computations, deep-learning methods provide promising solutions to the accuracy-efficiency dilemma inherent in DFT, enabling efficient derivation of physical quantities with ab initio-level accuracy for exascale materials. The order-of-magnitude reduction in computational time and the impressive transferability of deep-learned material models indicate a bright future for this field.

\begin{center}
\begin{figure}
    \centering
    \includegraphics[width=\linewidth]{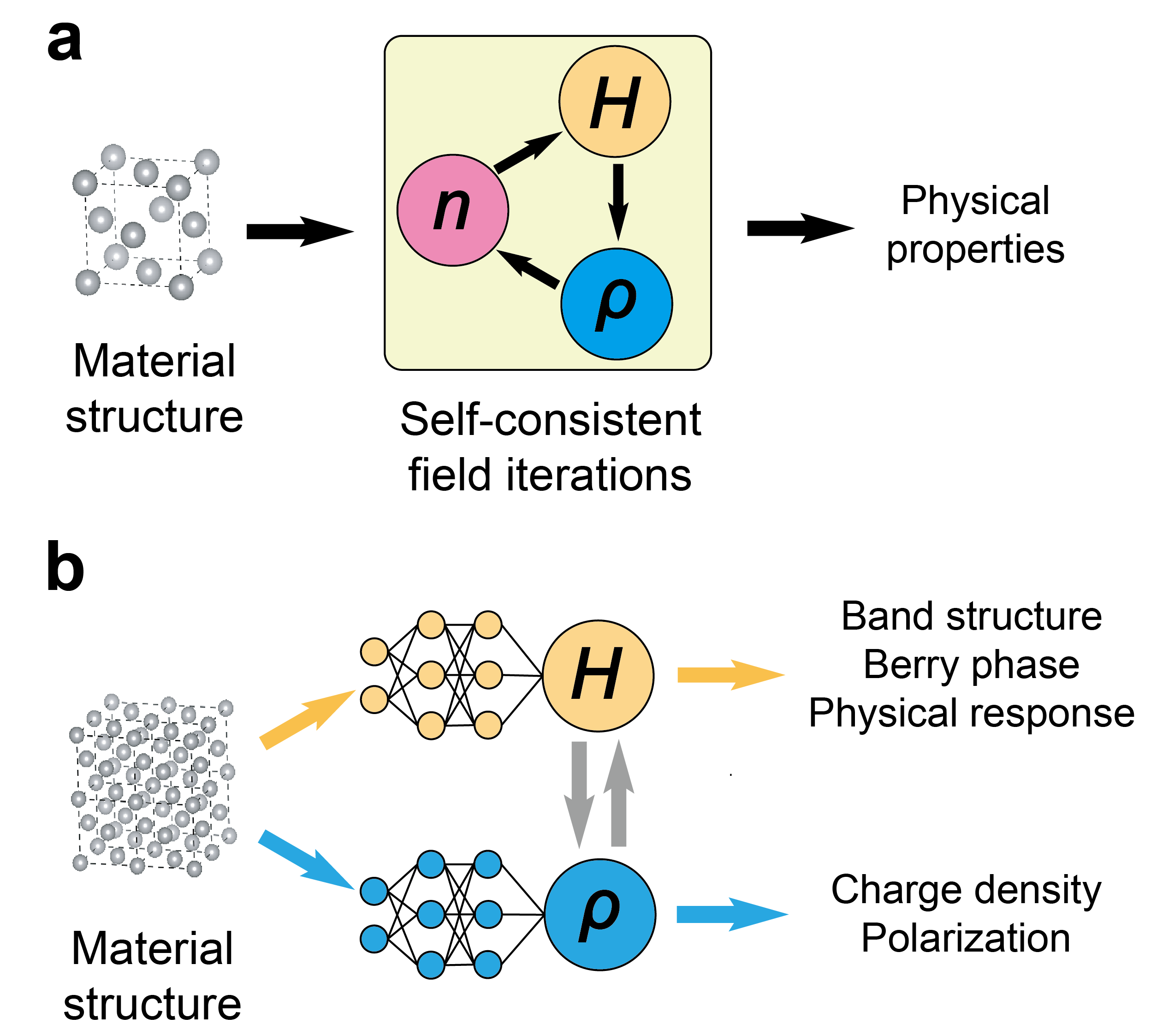}
    \caption{Schematic workflow of deep-learning DFT calculations, including DeepH-DM. \textbf{a} Traditional DFT simulation workflow. Starting from an input material structure and an initial guess, DFT programs perform self-consistent field (SCF) iterations involving three fundamental quantites: the DFT Hamiltonian $H$, the density matrix $\rho$ and the charge density $n$. Physical properties are derived after the SCF iterations converge. \textbf{b} Deep-learning DFT calculation workflow. Neural networks are trained to represent the $H$ or $\rho$ of materials structures, and can be used to predict $H$ or $\rho$ of unseen structures. While DeepH methods focus on predicting $H$, DeepH-DM models $\rho$ with neural networks. Various physical properties can be derived from the predictions made by DeepH or DeepH-DM.}
    \label{fig1}
\end{figure} 
\end{center}

Deep-learning electronic structure properties have gained significant attention. Unlike neural-network force fields, deep-learning electronic structure methods aim to predict the electronic properties, facilitating the prediction of band structures, electronic transportation properties, etc. Three fundamental quantities are involved in Kohn-Sham DFT for representing electronic structure, including DFT Hamiltonian $H$, one-body reduced density matrix (referred to as ``density matrix") $\rho$ and charge density $n$ (Fig.~\ref{fig1}\textbf{a}). Upon achieving self-consistency, each of these quantities can be used to compute the other two, indicating that all three contain sufficient information about the ground state electronic structure, and could therefore be regarded as ``fundamental" quantities in DFT electronic structures. While efforts have been made to seek the neural-network modeling of $H$~\cite{deeph2022,deeph-e32023,xdeeph2023,deeph-dfpt2024,deeph-hybrid2023,deeph22024,deephumm2024} or $n$~\cite{fabrizio2019electron,qiao2022informing}, the neural-network representation of $\rho$ remains elusive to date~\cite{shao2023machine}.

In this work, we introduce a neural-network method named DeepH-DM to establish the deep-learning of the density matrix in DFT programs using atomic-like localized bases. By leveraging the physical principles of ``quantum nearsightedness"~\cite{Kohn1996,Prodan2005} and fundamental physical equivariance, we demonstrated that the density matrix can be treated within the same framework as the Hamiltonian matrix. The deep-learning Hamiltonian architecture, DeepH-2~\cite{deeph22024}, is then adopted to represent the density matrix. Through example studies on test systems, we showcase the high accuracy and generalizability of the DeepH-DM method. Notably, the density matrix may be utilized for accurate predictions of charge density and for reproducing the Hamiltonian with a single step of non-self-consistent DFT computation.

Despite $\rho, n$ and $H$ all being fundamental quantities representing DFT electronic structures, the data size and the complexity of deriving other physical quantities from these quantities differ significantly. In DFT programs using atomic-like localized basis such as pseudo-atomic bases (PAO) as basis sets~\cite{siesta2002,openmx2004}, $\rho$ and $H$ are expressed in matrix forms $\rho_{\alpha\beta}$ and $H_{\alpha\beta}$, where $\alpha$ and $\beta$ denote indices of PAOs. In contrast, the charge density $n(\mathbf{r})$ is stored on a dense real-space grid, resulting in significantly larger data storage requirements compared to $\rho_{\alpha\beta}$ and $H_{\alpha\beta}$. Meanwhile, $n(\mathbf{r})$ can be derived by $\rho_{\alpha\beta}$ with the following equation:

\begin{align}
\label{rho}
n(\mathbf{r}) =\sum_{\alpha\beta} \rho_{\alpha\beta}\phi_\alpha^*(\mathbf{r})\phi_\beta(\mathbf{r})
\end{align}

Thus, $\rho_{\alpha\beta}$ can be considered a sparse representation of $n(\mathbf{r})$, and predicting $\rho$ instead of $n$ provides a more efficient method for modeling the electronic structure properties. 

Although $H_{\alpha\beta}$ and $\rho_{\alpha\beta}$ both provide sufficient information about the electronic structure, the computational cost for deriving specific quantities from them may vary. As depicted in Fig.~\ref{fig1}\textbf{b}, $H$ could be efficiently utilized for deriving the band structure near the Fermi surface, band topology properties such as Berry phase, and other physical responses, as demonstrated in our former works~\cite{deeph2022}. On the other hand, $\rho$ provides a more efficient derivation of $n$ or polarization. While such quantities are also theoretically accessible from $H$, the computation could be expensive, since derivation of $\rho$ from $H$ requires accessing the Kohn-Sham wavefunction of all occupied states, resulting in a computational complexity of $O(N^3)$ with respect to system size~\cite{Martin2004}. Thereby, the neural-network predictions of $H$ and $\rho$ are both of significance for physical properties computations.

Beyond the Kohn-Sham DFT, the one-body reduced density matrix also serves as a fundamental quantity in various calculation methods, including hybrid DFT functionals, density matrix functional theory, density matrix embedding theory and the construction of localized Wannier functions, etc.~\cite{deeph-hybrid2023,gilbert1975hohenberg,knizia2012density}. Therefore, DeepH-DM also bears broad potential application beyond the current DFT framework.

\section{Methods}

\subsection{Nearsightedness and sparsity}
In the realm of neural network modeling of physical quantities, two fundamental principles—nearsightedness and equivariance—play pivotal roles. From the ``quantum nearsightedness principle"~\cite{Kohn1996}, the nearsightedness of the real-space density operator $\rho(\textbf{r},\textbf{r}')$ plays as the cornerstone for the nearsightedness of all relevant physical quantities. Two properties present for $\rho(\textbf{r},\textbf{r}')$: 1. It decays significantly with respect to $\left|\textbf{r}'-\textbf{r}\right|$~\cite{Martin2004} 2. $\rho(\textbf{r},\textbf{r}')$ is only relevant to atoms within some nearsightedness range $R_{\text{N}}$~\cite{Prodan2005}. The density matrix $\rho_{\alpha\beta}$ denotes the representation of $\rho(\textbf{r},\textbf{r}')$ under localized bases, and therefore is also compatible with the nearsightedness principle. Message-passing graph neural networks (such as the DeepH-2 architecture) which determines local quantities based on nearby atomic structures is therefore suitable for DeepH-DM.

Due to the decaying tail of $\rho(\textbf{r},\textbf{r}')$, the matrix form of $\rho_{\alpha\beta}$ is not strictly sparse. However, only a sparse portion of $\rho_{\alpha\beta}$ needs to be predicted to determine the entire electronic structure. Within the Kohn-Sham framework, the electronic Hamiltonian is a functional of the charge density $n$, taking the form $H[n(\textbf{r})]$. From Eq.~\ref{rho}, only $\rho_{\alpha\beta}$ with overlapping $\phi_\alpha$ and $\phi_\beta$ contribute to $n(\textbf{r})$. For PAOs, $\phi_\alpha$ is the product of a spherical harmonic and an atom-centered localized radial function, resulting in a finite cutoff radius for the bases. Only $\rho_{\alpha\beta}$ within a finite distance needs to be predicted to accurately represent $n(\mathbf{r})$, and thus uniquely determine $H[n(\textbf{r})]$. Notably, the sparsity pattern of the density matrix is analogous to that of the Hamiltonian, since $H_{\alpha\beta}=\int d\textbf{r} \phi_{\alpha}^*(\textbf{r}) (-\frac{\nabla^2}{2m}+V_{\text{KS}}(\textbf{r}))\phi_{\beta}(\textbf{r})$ is non-zero only for overlapping $\phi_{\alpha}$ and $\phi_{\beta}$~\cite{siesta2002}. Therefore, the DeepH method could be naturally generalized to predicting the density matrix, giving rise to the DeepH-DM method.

\subsection{Equivariance of the density matrix}

Followingly, we demonstrate that the density matrix under localized bases is composed of $\text{E}(3)$ equivariant tensors. $\text{E}(3)$ represents the Euclidean group in three-dimensional space, encompassing spatial translation, inversion, and $\text{SO}(3)$ rotation. All PAOs are atom-centered, thereby ensuring the equivariance of $\rho_{\alpha\beta}$ with respect to spatial translation. To prove the equivariance with respect to spatial rotation, we denote the bases as $\phi_{iplm}=R_{ipl}(|r|)Y_{lm}(\hat{r})$, in which $i, p, l, m$ represents the atom index, multiplicity index, angular momentum and magnetic momentum index of a basis, respectively, and $|r|, \hat{r}$ represents the relative distance and direction of a grid point to the $i$-th atom nucleus. To discuss the transformation of bases under spatial rotation, each set of $(2l+1)$ bases with the same $ipl$ can be grouped together, denoted as $\left\{\phi_n\right\}\equiv\left\{\phi_{ipl}\right\}=\left(\phi_{ipl,-m},\phi_{ipl,-m+1},\cdots,\phi_{ipl,m}\right)^T$. This $(2l+1)$-dimensional vector carries an irreducible representation of $\text{SO}(3)$ of angular momentum $l$, transforming as $\left\{\phi_n\right\}\to D^{l_n}(\textbf{R})\left\{\phi_n\right\}$ upon a spatial rotation $\textbf{R}\in\text{SO}(3)$, where $D^l(\textbf{R})$ denotes the Wigner-D matrix. Thus, the transformation of the whole basis set is expressed as:

\begin{align}
\label{basis_transformation}
\begin{gathered}
\left\{\phi_1\right\}\oplus\left\{\phi_2\right\}\oplus\cdots\oplus\left\{\phi_n\right\}\to \\ \begin{pmatrix}D^{l_1}(\textbf{R})& & & \\ &D^{l_2}(\textbf{R}) & & \\ & &\ddots & \\ & & &D^{l_n}(\textbf{R})\end{pmatrix}\left\{\phi_1\right\}\oplus\left\{\phi_2\right\}\oplus\cdots\oplus\left\{\phi_n\right\}
\end{gathered}
\end{align}

Here, $\oplus$ represents direct sum between vectors. Below we abbreviate the transformation matrix in Eq.~\ref{basis_transformation} as $D^L(\textbf{R})$ and denote $\left\{\phi\right\}\equiv \left\{\phi_1\right\}\oplus\left\{\phi_2\right\}\oplus\cdots\oplus\left\{\phi_n\right\}$.

The equivariance of density matrix $\rho$ with respect to spatial rotation may be viewed with assistance of the ``dual basis"~\cite{ozaki2001convergent}. Dual basis is defined with $\{\tilde{\phi}\}=S^{-1}\cdot\left\{\phi\right\}$, in which $S^{-1}$ is the matrix inverse of the overlap matrix $S_{\alpha\beta}\equiv\int d\textbf{r}\phi_{\alpha}^*(\textbf{r})\phi_{\beta}(\textbf{r})$. It can be shown that $\int d\textbf{r}\tilde{\phi}_{\alpha}^*(\textbf{r})\phi_{\beta}(\textbf{r})=\delta_{\alpha\beta}$, where $\delta_{\alpha\beta}$ denotes the Kronecker delta. 

Recalling the basis transformation of the density operator $\rho$ between $\rho(\textbf{r},\textbf{r}')$ and $\rho_{\alpha\beta}$:

\begin{align}
\label{rho_r_to_basis}
\rho\left(\textbf{r},\textbf{r}'\right)=\sum_{\alpha\beta}\rho_{\alpha\beta}\phi_{\alpha}^*\left(\mathbf{r}\right)\phi_{\beta}\left(\mathbf{r}'\right)
\end{align}

The inverse transformation relationship writes:

\begin{align}
\label{rho_basis_to_r}
\rho_{\alpha\beta}=\int d\mathbf{r}\int d\mathbf{r}' \rho\left(\mathbf{r},\mathbf{r}'\right)\tilde{\phi}_{\alpha}^*\left(\mathbf{r}\right)\tilde{\phi}_{\beta}\left(\mathbf{r}'\right)
\end{align}

Upon rotation $\textbf{R}$, $\rho(\mathbf{r},\mathbf{r'})$ is invariant, since the coordinates $\mathbf{r},\mathbf{r'}$ rotates in the same way with the materials structure. Therefore the only quantity to be concerned is $\tilde{\phi}$. Based on the basis transformation relationship in Eq.~\ref{basis_transformation} and definition of dual basis, the transformation relationships of $S,S^{-1}$ and $\tilde{\phi}$ writes as follows:

\begin{align}
\begin{gathered}
\label{dual_basis_transformation}
S \to  D^L\left(\mathbf{R}\right) S \left(D^L\left(\mathbf{R}\right)\right)^\dagger \\
S^{-1} \to  D^L\left(\mathbf{R}\right) S^{-1} \left(D^L\left(\mathbf{R}\right)\right)^\dagger \\
\{\tilde{\phi}\}\to D^L\left(\textbf{R}\right)\{\tilde{\phi}\}
\end{gathered}
\end{align}

Combining Eq.~\ref{rho_basis_to_r} and Eq.~\ref{dual_basis_transformation}, we have the density matrix $\left[\rho_{\alpha\beta}\right]$ transforms as:

\begin{align}
\label{density_matrix_transformation}
\left[\rho_{\alpha\beta}\right]\to D^L\left(\mathbf{R}\right) \left[\rho_{\alpha\beta}\right] \left(D^L\left(\mathbf{R}\right)\right)^\dagger
\end{align}

Therefore, $\rho_{\alpha\beta}$ is proved to be an $\text{SO}(3)$ equivariant tensor carrying representation $(l_1\oplus l_2\oplus\cdots\oplus l_n)\otimes(l_1\oplus l_2\oplus\cdots\oplus l_n)$, in which $\oplus$ is the direct product of groups. The sparse form of density matrix can also be divided into equivariant sub-blocks carrying representation $l_{i}\otimes l_{j}$, which can be treated within the same framework with the Hamiltonians.

\subsection{Neural network framework}
The neural network framework applied in this work is DeepH-2~\cite{deeph22024}, an equivariant local coordinate transformer inspired by eSCN and EquiformerV2~\cite{escn2023,liao2024equiformerv2}. Compared to the our former realization of equivariant neural network prediction of Hamiltonian (DeepH-E3)~\cite{deeph-e32023}, DeepH-2 features incorporation of attention mechanism, together with the $\text{SO}(2)$ tensor product implemented in eSCN~\cite{escn2023}. In analogy with DeepH-E3, DeepH-2 is an equivariant graph neural network. From the input material structure, atoms and atom pairs are mapped to the node and edge features of the neural network, respectively. The input structure information, including information for atomic position and atomic species, is embedded into equivariant feature vectors. Each feature vector carries an irreducible representation of the $\text{SO}(3)$ group labelled by angular momentum $l$. All intermediate neural network operations are constrained to preserve the equivariance of neural network features. The DeepH-2 consists multiple message-passing blocks, in which node and edge features are updated with nearby features. Information for atomic structure thus gradually concentrate to distant atoms through message-passing blocks. As proved in the above section, the density matrix is composed of $\text{SO}(3)$ equivariant tensors, analogous to the Hamiltonian. The Wigner-Eckart theorem is finally applied to construct the equivariant tensor from equivariant feature vectors~\cite{deeph-e32023}.

\begin{center}
\begin{figure*}
    \centering
    \includegraphics[width=\linewidth]{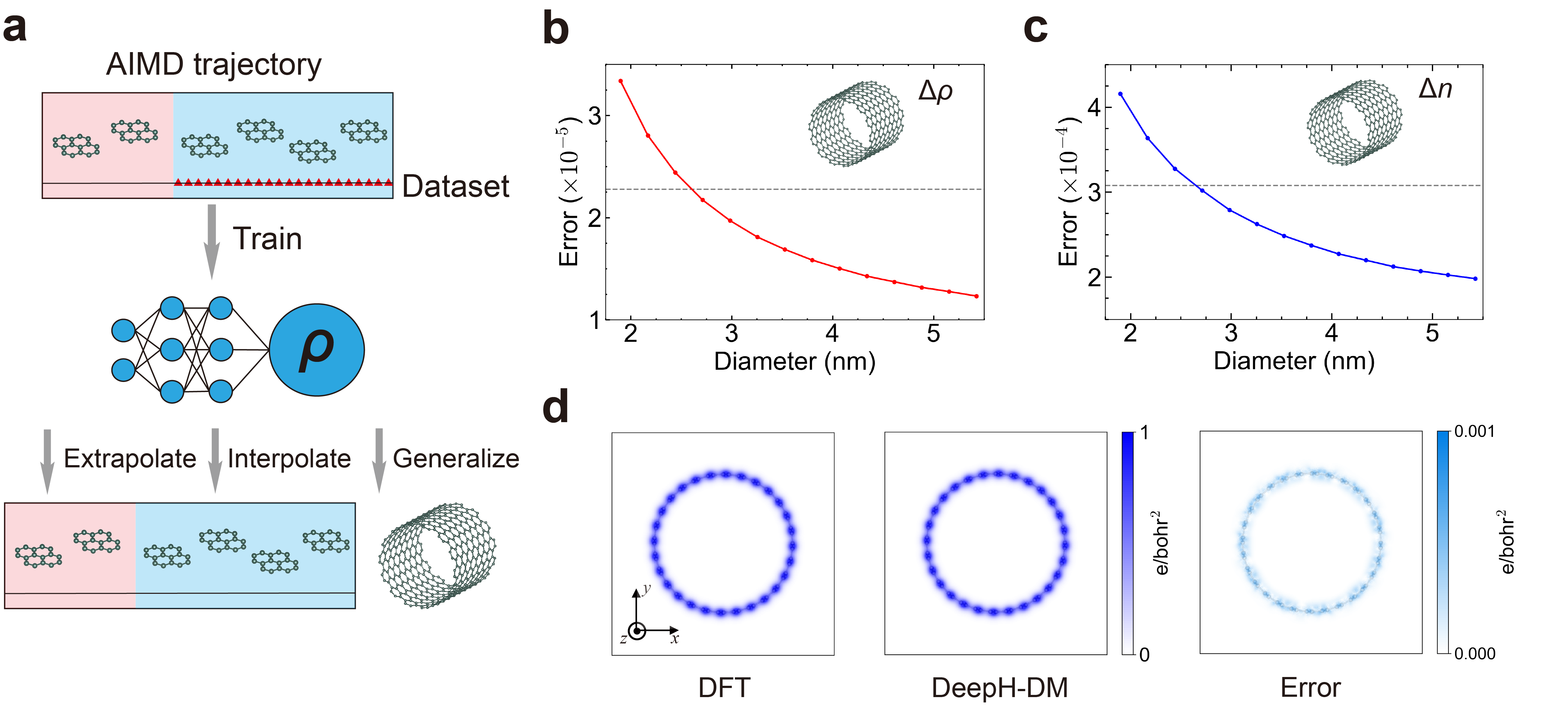}
    \caption{Performance of DeepH-DM on monolayer graphene. \textbf{a} Schematic workflow. Datasets of monolayer graphene are sampled from ab initio molecular dynamics (AIMD) trajectory. The trained DeepH-DM model can be used for extrapolation or interpolation on AIMD trajectory structures and can be generalized to carbon nanotubes (CNTs). \textbf{b,c} DeepH-DM's performance on CNTs. Error of DeepH-DM on CNTs with different diameters, including error for the density matrix $\Delta\rho$ (\textbf{b}) and for charge density $\Delta n$ (\textbf{c}). The test-set mean error of the monolayer graphene dataset are depicted with dashed lines. \textbf{d} DFT-calculated, DeepH-DM-predicted, and absolute error of charge density of the CNT with diameter $1.90$~nm (with chiral index $(14, 14)$). The charge density is integrated over the out-of-plane direction (the $z$-axis). The error in the charge density is shown on a colorbar three orders of magnitude smaller.}
    \label{fig2}
\end{figure*} 
\end{center}

\subsection{Dataset preparation}
Two datasets are present for case studies, including monolayer and bilayer graphene, with the same structure as in Ref.~\cite{deeph-e32023}. In specific, structures are sampled from ab initio molecular dynamic trajectory of $6\times6$ supercells for monolayer graphene (including 450 structures) and $4\times4$ supercells for bilayer graphene (including 300 structures). The dataset is computed with the OpenMX package~\cite{openmx2003,openmx2004}, applying norm-conserving pseudopotential~\cite{ncpp1979} and Perdew-Berke-Ernzerhof exchange-correlation functional~\cite{pbe1996}. The ``standard"-size basis set defined by OpenMX is applied for both dataset, with 13 orbitals for each carbon atom and a cutoff range of $6.0$~bohr. To guarantee the convergence of the density matrix, a dense Monkhorst-Pack $\textbf{k}$-mesh is applied to each dataset, namely, $5\times5\times1$ for both monolayer and bilayer graphene dataset.

\section{Results}

The capability of DeepH-DM is first demonstrated using the monolayer graphene dataset, as depicted in Fig.~\ref{fig2}\textbf{a}. A straightforward application of DeepH-DM is to compute the charge density $n$ from the predicted density matrix $\rho$ following Eq.~\ref{rho}. To assess the accuracy of DeepH-DM model, we calculate the mean absolute error (MAE) of density matrix (denoted $\Delta\rho$) and the relative error of the charge density, defined as: 

\begin{align}
\label{def_Deltarho}
\frac{\int d\mathbf{r}|n_{\text{DeepH}}(\mathbf{r})-n_{\text{DFT}}(\mathbf{r})|}{\int d\mathbf{r} n_{\text{DFT}}(\mathbf{r})}
\end{align}

\begin{center}
\begin{figure*}
    \centering
    \includegraphics[width=\linewidth]{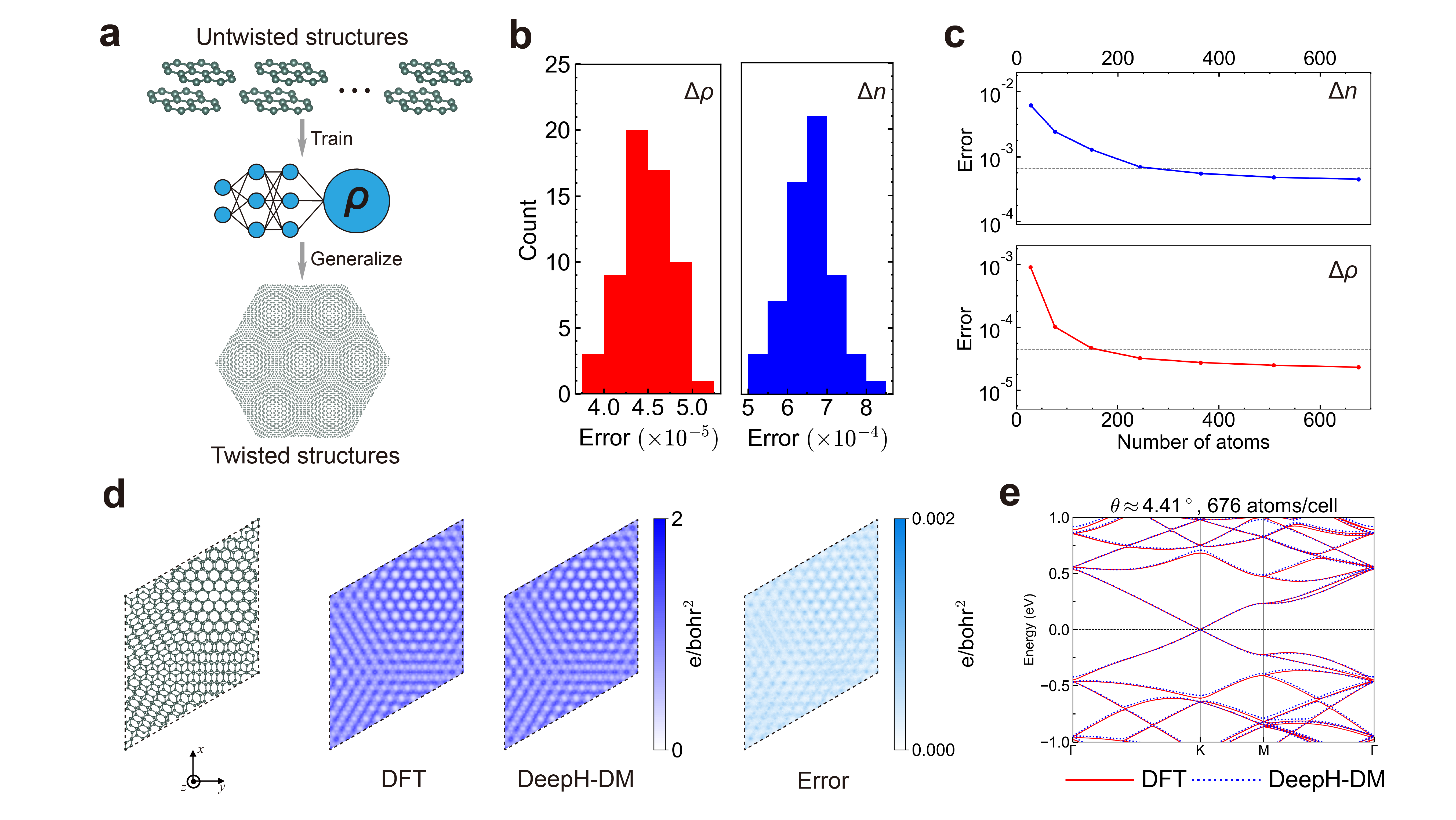}
    \caption{Performance of DeepH-DM on bilayer graphene. \textbf{a} Schematic workflow. A DeepH-DM model is trained on a dataset comprising untwisted bilayer graphene structures and is capable of generalizing to twisted bilayer graphenes (TBGs). \textbf{b} Distribution of error of density matrix $\Delta\rho$ and charge density $\Delta n$ of DeepH-DM model on test-set structures. \textbf{c} Error of DeepH-DM on TBGs. The error of density matrix and charge density are evaluated on $(m, m-1)$ TBGs, with $m$ ranging from 2 to 8. \textbf{d} DFT-calculated, DeepH-DM-predicted, and absolute error in charge density of TBGs with twist index $(8, 7)$. The charge density is integrated over the out-of-plane direction (the $z$-axis). \textbf{e} Comparison of DFT-calculated and DeepH-DM-predicted band structures of the $(8, 7)$ TBG.}
    \label{fig3}
\end{figure*} 
\end{center}

Here, $n_{\text{DFT}}$ and $n_{\text{DeepH}}$ denote charge density computed from DFT and the predicted density matrix, respectively. The 450 structures involved in the training process are sampled from 5,000 structures in an ab-initio molecular dynamics trajectory. Specifically, the 450 structures are sampled every 10 frames within the last 4,500 frames. Thereby, we can divide the 5,000 structures into three subsets: 450 structures which comprise the last dataset used in the DeepH-DM training process (referred to as ``dataset structures"); 4,050 structures within the 4,500 frames but unused in training (referred to as ``interpolation structures"); and 500 structures outside the sampled region (referred to as ``extrapolation structures"). The ``extrapolation structures" may be used to measure the methods' generalizability. The overall averaged $\Delta \rho$ across interpolation and extrapolation structures are $2.21\times10^{-5}$ and $2.24\times10^{-5}$, respectively, both of which are comparable to the test-set MAE ($2.28\times10^{-5}$). The overall averaged $\Delta n$ across interpolation and extrapolation structures are $2.93\times10^{-4}$ and $2.94\times10^{-4}$, respectively, also comparable to the test-set MAE of $3.08\times10^{-4}$. 

In addition, the DeepH-DM model trained on monolayer graphene demonstrates its capability to generalize to carbon nanotubes (CNTs). This capability is examined on a series of armchair CNTs. The relationship between $\Delta\rho, \Delta n$, and the diameter of the CNTs is depicted in Fig.~\ref{fig2}\textbf{b,c}, respectively, with dashed lines denoting the mean error of the test structures of monolayer graphene. The DeepH-DM model shows a decreasing trend in error with increasing diameter, attributed to the decreased curvature of CNTs with larger diameters. For CNTs with diameters greater than $2.5$~nm, the error is even smaller than that of the test set for monolayer graphene, which can be attributed to the high symmetry of CNTs. These symmetries are fully exploited by the equivariant neural network architecture of DeepH-DM, simplifying the predictions. The charge density from DFT, DeepH-DM and their error for the $(14, 14)$ CNT (corresponding to the leftmost point in Fig.~\ref{fig2}\textbf{b,c}) are shown in Fig.~\ref{fig2}\textbf{d}. Here, we plot the integrated density along the out-of-plane direction $z$ (for instance, $\int dz |n_{\text{DeepH}}(\textbf{r})-n_{\text{DFT}}(\textbf{r})|$ is plotted for error of the charge density). The error of the charge density has a colorbar three orders of magnitude smaller compared with the colorbar of the original charge density.

The DeepH-DM method is also applicable to bilayer graphene, including twisted bilayer graphenes (TBGs). Following the strategy in our former works~\cite{deeph2022}, DeepH-DM model is trained on the untwisted bilayer graphene structures with random inter-layer shift, and can generalize to twisted bilayers (Fig.~\ref{fig3}\textbf{a}). The error distribution of $\Delta\rho$ and $\Delta n$ of test-set structures are summarized in Fig.~\ref{fig3}\textbf{b}, both showing a narrow distribution of errors. The MAE for density matrix is $4.48\times 10^{-5}$, while the relative error of charge density averaged accross structures is $6.58\times10^{-4}$. Notably, even the worst test structure encompasses a relative error of $\Delta n<0.1\%$. 

As an important application, the DeepH-DM method is further applied to TBGs with twist indices $(m, m-1)$ ranging from $m=2$ to $m=8$, with results summarized in Fig.~\ref{fig3}\textbf{c}. The leftmost data points correspond to TBGs with larger twist angles and smaller Moir\'{e} supercells. As the twist angle decreases, the prediction error also decreases to values smaller than test-set average of untwisted structures (depicted by dashed lines). The charge distribution of a representative TBG (with twist indices $(8, 7)$, corresponding to 676 atoms per Moir\'{e} supercell) is illustrated in Fig.~\ref{fig3}\textbf{d}. Notably, the colorbar for the error of the charge density is three orders of magnitudes smaller. Band structures computed from the Hamiltonians reconstructed from the predicted density matrices are compared with DFT band structures in Fig.~\ref{fig3}\textbf{e}, showing a precise matching.

\section{Conclusion}

In summary, we introduced a deep-learning framework named DeepH-DM in pursuit of the neural-network modelling of the DFT density matrices. By analyzing the properties of the density matrix, including nearsightedness, sparsity, and equivariance, we demonstrated the deep learning of the density matrix can be treated in a unified framework with deep-learning Hamiltonians. We then applied the DeepH-2 architecture to deep-learning density matrix, showcasing its capabilities through example studies on monolayer and bilayer graphene. Given that both the density matrix and Hamiltonian are fundamental quantities in DFT calculations, this work establishes a deep-learning framework that is as crucial to electronic structure calculations as the deep-learning Hamiltonian methods.

\section{acknowledgments}
This work was supported by the Basic Science Center Project of NSFC (grant no. 52388201), the National Natural Science Foundation of China (grant no. 12334003), the National Science Fund for Distinguished Young Scholars (grant no. 12025405), the Ministry of Science and Technology of China (grant no. 2023YFA1406400), the Beijing Advanced Innovation Center for Future Chip (ICFC), and the Beijing Advanced Innovation Center for Materials Genome Engineering. The work was carried out at National Supercomputer Center in Tianjin using the Tianhe new generation supercomputer.


\begin{thebibliography}{29}%
    \makeatletter
    \providecommand \@ifxundefined [1]{%
     \@ifx{#1\undefined}
    }%
    \providecommand \@ifnum [1]{%
     \ifnum #1\expandafter \@firstoftwo
     \else \expandafter \@secondoftwo
     \fi
    }%
    \providecommand \@ifx [1]{%
     \ifx #1\expandafter \@firstoftwo
     \else \expandafter \@secondoftwo
     \fi
    }%
    \providecommand \natexlab [1]{#1}%
    \providecommand \enquote  [1]{``#1''}%
    \providecommand \bibnamefont  [1]{#1}%
    \providecommand \bibfnamefont [1]{#1}%
    \providecommand \citenamefont [1]{#1}%
    \providecommand \href@noop [0]{\@secondoftwo}%
    \providecommand \href [0]{\begingroup \@sanitize@url \@href}%
    \providecommand \@href[1]{\@@startlink{#1}\@@href}%
    \providecommand \@@href[1]{\endgroup#1\@@endlink}%
    \providecommand \@sanitize@url [0]{\catcode `\\12\catcode `\$12\catcode
      `\&12\catcode `\#12\catcode `\^12\catcode `\_12\catcode `\%12\relax}%
    \providecommand \@@startlink[1]{}%
    \providecommand \@@endlink[0]{}%
    \providecommand \url  [0]{\begingroup\@sanitize@url \@url }%
    \providecommand \@url [1]{\endgroup\@href {#1}{\urlprefix }}%
    \providecommand \urlprefix  [0]{URL }%
    \providecommand \Eprint [0]{\href }%
    \providecommand \doibase [0]{https://doi.org/}%
    \providecommand \selectlanguage [0]{\@gobble}%
    \providecommand \bibinfo  [0]{\@secondoftwo}%
    \providecommand \bibfield  [0]{\@secondoftwo}%
    \providecommand \translation [1]{[#1]}%
    \providecommand \BibitemOpen [0]{}%
    \providecommand \bibitemStop [0]{}%
    \providecommand \bibitemNoStop [0]{.\EOS\space}%
    \providecommand \EOS [0]{\spacefactor3000\relax}%
    \providecommand \BibitemShut  [1]{\csname bibitem#1\endcsname}%
    \let\auto@bib@innerbib\@empty
    \bibitem [{\citenamefont {Behler}\ and\ \citenamefont
      {Parrinello}(2007)}]{Behler2007}%
      \BibitemOpen
      \bibfield  {author} {\bibinfo {author} {\bibfnamefont {J.}~\bibnamefont
      {Behler}}\ and\ \bibinfo {author} {\bibfnamefont {M.}~\bibnamefont
      {Parrinello}},\ }\bibfield  {title} {\bibinfo {title} {Generalized
      neural-network representation of high-dimensional potential-energy
      surfaces},\ }\href {https://doi.org/10.1103/PhysRevLett.98.146401} {\bibfield
       {journal} {\bibinfo  {journal} {Phys. Rev. Lett.}\ }\textbf {\bibinfo
      {volume} {98}},\ \bibinfo {pages} {146401} (\bibinfo {year}
      {2007})}\BibitemShut {NoStop}%
    \bibitem [{\citenamefont {Zhang}\ \emph {et~al.}(2018)\citenamefont {Zhang},
      \citenamefont {Han}, \citenamefont {Wang}, \citenamefont {Car},\ and\
      \citenamefont {E}}]{Zhang2018}%
      \BibitemOpen
      \bibfield  {author} {\bibinfo {author} {\bibfnamefont {L.}~\bibnamefont
      {Zhang}}, \bibinfo {author} {\bibfnamefont {J.}~\bibnamefont {Han}}, \bibinfo
      {author} {\bibfnamefont {H.}~\bibnamefont {Wang}}, \bibinfo {author}
      {\bibfnamefont {R.}~\bibnamefont {Car}},\ and\ \bibinfo {author}
      {\bibfnamefont {W.}~\bibnamefont {E}},\ }\bibfield  {title} {\bibinfo {title}
      {Deep potential molecular dynamics: A scalable model with the accuracy of
      quantum mechanics},\ }\href {https://doi.org/10.1103/PhysRevLett.120.143001}
      {\bibfield  {journal} {\bibinfo  {journal} {Phys. Rev. Lett.}\ }\textbf
      {\bibinfo {volume} {120}},\ \bibinfo {pages} {143001} (\bibinfo {year}
      {2018})}\BibitemShut {NoStop}%
    \bibitem [{\citenamefont {Sch\"{u}tt}\ \emph {et~al.}(2019)\citenamefont
      {Sch\"{u}tt}, \citenamefont {Gastegger}, \citenamefont {Tkatchenko},
      \citenamefont {M\"{u}ller},\ and\ \citenamefont {Maurer}}]{Schutt2019}%
      \BibitemOpen
      \bibfield  {author} {\bibinfo {author} {\bibfnamefont {K.~T.}\ \bibnamefont
      {Sch\"{u}tt}}, \bibinfo {author} {\bibfnamefont {M.}~\bibnamefont
      {Gastegger}}, \bibinfo {author} {\bibfnamefont {A.}~\bibnamefont
      {Tkatchenko}}, \bibinfo {author} {\bibfnamefont {K.-R.}\ \bibnamefont
      {M\"{u}ller}},\ and\ \bibinfo {author} {\bibfnamefont {R.~J.}\ \bibnamefont
      {Maurer}},\ }\bibfield  {title} {\bibinfo {title} {Unifying machine learning
      and quantum chemistry with a deep neural network for molecular
      wavefunctions},\ }\href {https://doi.org/10.1038/s41467-019-12875-2}
      {\bibfield  {journal} {\bibinfo  {journal} {Nat. Commun.}\ }\textbf {\bibinfo
      {volume} {10}},\ \bibinfo {pages} {5024} (\bibinfo {year}
      {2019})}\BibitemShut {NoStop}%
    \bibitem [{\citenamefont {Unke}\ \emph {et~al.}(2021)\citenamefont {Unke},
      \citenamefont {Bogojeski}, \citenamefont {Gastegger}, \citenamefont {Geiger},
      \citenamefont {Smidt},\ and\ \citenamefont {M\"{u}ller}}]{Unke2021}%
      \BibitemOpen
      \bibfield  {author} {\bibinfo {author} {\bibfnamefont {O.~T.}\ \bibnamefont
      {Unke}}, \bibinfo {author} {\bibfnamefont {M.}~\bibnamefont {Bogojeski}},
      \bibinfo {author} {\bibfnamefont {M.}~\bibnamefont {Gastegger}}, \bibinfo
      {author} {\bibfnamefont {M.}~\bibnamefont {Geiger}}, \bibinfo {author}
      {\bibfnamefont {T.}~\bibnamefont {Smidt}},\ and\ \bibinfo {author}
      {\bibfnamefont {K.-R.}\ \bibnamefont {M\"{u}ller}},\ }\bibfield  {title}
      {\bibinfo {title} {{SE}(3)-equivariant prediction of molecular wavefunctions
      and electronic densities},\ }in\ \href
      {https://openreview.net/forum?id=auGY2UQfhSu} {\emph {\bibinfo {booktitle}
      {Advances in Neural Information Processing Systems}}}\ (\bibinfo  {publisher}
      {Curran Associates, Inc.},\ \bibinfo {year} {2021})\ p.\ \bibinfo {pages}
      {14434}\BibitemShut {NoStop}%
    \bibitem [{\citenamefont {Li}\ \emph {et~al.}(2022)\citenamefont {Li},
      \citenamefont {Wang}, \citenamefont {Zou}, \citenamefont {Ye}, \citenamefont
      {Xu}, \citenamefont {Gong}, \citenamefont {Duan},\ and\ \citenamefont
      {Xu}}]{deeph2022}%
      \BibitemOpen
      \bibfield  {author} {\bibinfo {author} {\bibfnamefont {H.}~\bibnamefont
      {Li}}, \bibinfo {author} {\bibfnamefont {Z.}~\bibnamefont {Wang}}, \bibinfo
      {author} {\bibfnamefont {N.}~\bibnamefont {Zou}}, \bibinfo {author}
      {\bibfnamefont {M.}~\bibnamefont {Ye}}, \bibinfo {author} {\bibfnamefont
      {R.}~\bibnamefont {Xu}}, \bibinfo {author} {\bibfnamefont {X.}~\bibnamefont
      {Gong}}, \bibinfo {author} {\bibfnamefont {W.}~\bibnamefont {Duan}},\ and\
      \bibinfo {author} {\bibfnamefont {Y.}~\bibnamefont {Xu}},\ }\bibfield
      {title} {\bibinfo {title} {Deep-learning density functional theory
      {Hamiltonian} for efficient ab initio electronic-structure calculation},\
      }\href {https://doi.org/10.1038/s43588-022-00265-6} {\bibfield  {journal}
      {\bibinfo  {journal} {Nat. Comput. Sci.}\ }\textbf {\bibinfo {volume} {2}},\
      \bibinfo {pages} {367} (\bibinfo {year} {2022})}\BibitemShut {NoStop}%
    \bibitem [{\citenamefont {Gong}\ \emph {et~al.}(2023)\citenamefont {Gong},
      \citenamefont {Li}, \citenamefont {Zou}, \citenamefont {Xu}, \citenamefont
      {Duan},\ and\ \citenamefont {Xu}}]{deeph-e32023}%
      \BibitemOpen
      \bibfield  {author} {\bibinfo {author} {\bibfnamefont {X.}~\bibnamefont
      {Gong}}, \bibinfo {author} {\bibfnamefont {H.}~\bibnamefont {Li}}, \bibinfo
      {author} {\bibfnamefont {N.}~\bibnamefont {Zou}}, \bibinfo {author}
      {\bibfnamefont {R.}~\bibnamefont {Xu}}, \bibinfo {author} {\bibfnamefont
      {W.}~\bibnamefont {Duan}},\ and\ \bibinfo {author} {\bibfnamefont
      {Y.}~\bibnamefont {Xu}},\ }\bibfield  {title} {\bibinfo {title} {General
      framework for {E}(3)-equivariant neural network representation of density
      functional theory {Hamiltonian}},\ }\href
      {https://doi.org/10.1038/s41467-023-38468-8} {\bibfield  {journal} {\bibinfo
      {journal} {Nat. Commun.}\ }\textbf {\bibinfo {volume} {14}},\ \bibinfo
      {pages} {2848} (\bibinfo {year} {2023})}\BibitemShut {NoStop}%
    \bibitem [{\citenamefont {Li}\ \emph {et~al.}(2023)\citenamefont {Li},
      \citenamefont {Tang}, \citenamefont {Gong}, \citenamefont {Zou},
      \citenamefont {Duan},\ and\ \citenamefont {Xu}}]{xdeeph2023}%
      \BibitemOpen
      \bibfield  {author} {\bibinfo {author} {\bibfnamefont {H.}~\bibnamefont
      {Li}}, \bibinfo {author} {\bibfnamefont {Z.}~\bibnamefont {Tang}}, \bibinfo
      {author} {\bibfnamefont {X.}~\bibnamefont {Gong}}, \bibinfo {author}
      {\bibfnamefont {N.}~\bibnamefont {Zou}}, \bibinfo {author} {\bibfnamefont
      {W.}~\bibnamefont {Duan}},\ and\ \bibinfo {author} {\bibfnamefont
      {Y.}~\bibnamefont {Xu}},\ }\bibfield  {title} {\bibinfo {title}
      {Deep-learning electronic-structure calculation of magnetic
      superstructures},\ }\href {https://doi.org/10.1038/s43588-023-00424-3}
      {\bibfield  {journal} {\bibinfo  {journal} {Nat. Comput. Sci.}\ }\textbf
      {\bibinfo {volume} {3}},\ \bibinfo {pages} {321} (\bibinfo {year}
      {2023})}\BibitemShut {NoStop}%
    \bibitem [{\citenamefont {Li}\ \emph {et~al.}(2024)\citenamefont {Li},
      \citenamefont {Tang}, \citenamefont {Fu}, \citenamefont {Dong}, \citenamefont
      {Zou}, \citenamefont {Gong}, \citenamefont {Duan},\ and\ \citenamefont
      {Xu}}]{deeph-dfpt2024}%
      \BibitemOpen
      \bibfield  {author} {\bibinfo {author} {\bibfnamefont {H.}~\bibnamefont
      {Li}}, \bibinfo {author} {\bibfnamefont {Z.}~\bibnamefont {Tang}}, \bibinfo
      {author} {\bibfnamefont {J.}~\bibnamefont {Fu}}, \bibinfo {author}
      {\bibfnamefont {W.-H.}\ \bibnamefont {Dong}}, \bibinfo {author}
      {\bibfnamefont {N.}~\bibnamefont {Zou}}, \bibinfo {author} {\bibfnamefont
      {X.}~\bibnamefont {Gong}}, \bibinfo {author} {\bibfnamefont {W.}~\bibnamefont
      {Duan}},\ and\ \bibinfo {author} {\bibfnamefont {Y.}~\bibnamefont {Xu}},\
      }\bibfield  {title} {\bibinfo {title} {Deep-learning density functional
      perturbation theory},\ }\href
      {https://doi.org/10.1103/PhysRevLett.132.096401} {\bibfield  {journal}
      {\bibinfo  {journal} {Phys. Rev. Lett.}\ }\textbf {\bibinfo {volume} {132}},\
      \bibinfo {pages} {096401} (\bibinfo {year} {2024})}\BibitemShut {NoStop}%
    \bibitem [{\citenamefont {Tang}\ \emph {et~al.}(2023)\citenamefont {Tang},
      \citenamefont {Li}, \citenamefont {Lin}, \citenamefont {Gong}, \citenamefont
      {Jin}, \citenamefont {He}, \citenamefont {Jiang}, \citenamefont {Ren},
      \citenamefont {Duan},\ and\ \citenamefont {Xu}}]{deeph-hybrid2023}%
      \BibitemOpen
      \bibfield  {author} {\bibinfo {author} {\bibfnamefont {Z.}~\bibnamefont
      {Tang}}, \bibinfo {author} {\bibfnamefont {H.}~\bibnamefont {Li}}, \bibinfo
      {author} {\bibfnamefont {P.}~\bibnamefont {Lin}}, \bibinfo {author}
      {\bibfnamefont {X.}~\bibnamefont {Gong}}, \bibinfo {author} {\bibfnamefont
      {G.}~\bibnamefont {Jin}}, \bibinfo {author} {\bibfnamefont {L.}~\bibnamefont
      {He}}, \bibinfo {author} {\bibfnamefont {H.}~\bibnamefont {Jiang}}, \bibinfo
      {author} {\bibfnamefont {X.}~\bibnamefont {Ren}}, \bibinfo {author}
      {\bibfnamefont {W.}~\bibnamefont {Duan}},\ and\ \bibinfo {author}
      {\bibfnamefont {Y.}~\bibnamefont {Xu}},\ }\bibfield  {title} {\bibinfo
      {title} {Efficient hybrid density functional calculation by deep learning},\
      }\href {https://doi.org/10.48550/arXiv.2302.08221} {\bibfield  {journal}
      {\bibinfo  {journal} {arXiv:2302.08221}\ } (\bibinfo {year}
      {2023})}\BibitemShut {NoStop}%
    \bibitem [{\citenamefont {Wang}\ \emph
      {et~al.}(2024{\natexlab{a}})\citenamefont {Wang}, \citenamefont {Li},
      \citenamefont {Tang}, \citenamefont {Tao}, \citenamefont {Wang},
      \citenamefont {Yuan}, \citenamefont {Chen}, \citenamefont {Duan},\ and\
      \citenamefont {Xu}}]{deeph22024}%
      \BibitemOpen
      \bibfield  {author} {\bibinfo {author} {\bibfnamefont {Y.}~\bibnamefont
      {Wang}}, \bibinfo {author} {\bibfnamefont {H.}~\bibnamefont {Li}}, \bibinfo
      {author} {\bibfnamefont {Z.}~\bibnamefont {Tang}}, \bibinfo {author}
      {\bibfnamefont {H.}~\bibnamefont {Tao}}, \bibinfo {author} {\bibfnamefont
      {Y.}~\bibnamefont {Wang}}, \bibinfo {author} {\bibfnamefont {Z.}~\bibnamefont
      {Yuan}}, \bibinfo {author} {\bibfnamefont {Z.}~\bibnamefont {Chen}}, \bibinfo
      {author} {\bibfnamefont {W.}~\bibnamefont {Duan}},\ and\ \bibinfo {author}
      {\bibfnamefont {Y.}~\bibnamefont {Xu}},\ }\bibfield  {title} {\bibinfo
      {title} {{DeepH-2}: Enhancing deep-learning electronic structure via an
      equivariant local-coordinate transformer},\ }\href
      {https://doi.org/10.48550/arXiv.2401.17015} {\bibfield  {journal} {\bibinfo
      {journal} {arXiv:2401.17015}\ } (\bibinfo {year}
      {2024}{\natexlab{a}})}\BibitemShut {NoStop}%
    \bibitem [{\citenamefont {Wang}\ \emph
      {et~al.}(2024{\natexlab{b}})\citenamefont {Wang}, \citenamefont {Li},
      \citenamefont {Tang}, \citenamefont {Li}, \citenamefont {Yuan}, \citenamefont
      {Tao}, \citenamefont {Zou}, \citenamefont {Bao}, \citenamefont {Liang},
      \citenamefont {Chen} \emph {et~al.}}]{deephumm2024}%
      \BibitemOpen
      \bibfield  {author} {\bibinfo {author} {\bibfnamefont {Y.}~\bibnamefont
      {Wang}}, \bibinfo {author} {\bibfnamefont {Y.}~\bibnamefont {Li}}, \bibinfo
      {author} {\bibfnamefont {Z.}~\bibnamefont {Tang}}, \bibinfo {author}
      {\bibfnamefont {H.}~\bibnamefont {Li}}, \bibinfo {author} {\bibfnamefont
      {Z.}~\bibnamefont {Yuan}}, \bibinfo {author} {\bibfnamefont {H.}~\bibnamefont
      {Tao}}, \bibinfo {author} {\bibfnamefont {N.}~\bibnamefont {Zou}}, \bibinfo
      {author} {\bibfnamefont {T.}~\bibnamefont {Bao}}, \bibinfo {author}
      {\bibfnamefont {X.}~\bibnamefont {Liang}}, \bibinfo {author} {\bibfnamefont
      {Z.}~\bibnamefont {Chen}}, \emph {et~al.},\ }\bibfield  {title} {\bibinfo
      {title} {Universal materials model of deep-learning density functional theory
      {Hamiltonian}},\ }\href
      {https://www.sciencedirect.com/science/article/pii/S2095927324004079}
      {\bibfield  {journal} {\bibinfo  {journal} {Sci. Bull.}\ } (\bibinfo {year}
      {2024}{\natexlab{b}})}\BibitemShut {NoStop}%
    \bibitem [{\citenamefont {Yuan}\ \emph {et~al.}(2024)\citenamefont {Yuan},
      \citenamefont {Xu}, \citenamefont {Li}, \citenamefont {Cheng}, \citenamefont
      {Tao}, \citenamefont {Tang}, \citenamefont {Zhou}, \citenamefont {Duan},\
      and\ \citenamefont {Xu}}]{magnet2024}%
      \BibitemOpen
      \bibfield  {author} {\bibinfo {author} {\bibfnamefont {Z.}~\bibnamefont
      {Yuan}}, \bibinfo {author} {\bibfnamefont {Z.}~\bibnamefont {Xu}}, \bibinfo
      {author} {\bibfnamefont {H.}~\bibnamefont {Li}}, \bibinfo {author}
      {\bibfnamefont {X.}~\bibnamefont {Cheng}}, \bibinfo {author} {\bibfnamefont
      {H.}~\bibnamefont {Tao}}, \bibinfo {author} {\bibfnamefont {Z.}~\bibnamefont
      {Tang}}, \bibinfo {author} {\bibfnamefont {Z.}~\bibnamefont {Zhou}}, \bibinfo
      {author} {\bibfnamefont {W.}~\bibnamefont {Duan}},\ and\ \bibinfo {author}
      {\bibfnamefont {Y.}~\bibnamefont {Xu}},\ }\bibfield  {title} {\bibinfo
      {title} {Equivariant neural network force fields for magnetic materials},\
      }\href {https://doi.org/10.1007/s44214-024-00055-3} {\bibfield  {journal}
      {\bibinfo  {journal} {Quantum Front.}\ }\textbf {\bibinfo {volume} {3}},\
      \bibinfo {pages} {8} (\bibinfo {year} {2024})}\BibitemShut {NoStop}%
    \bibitem [{\citenamefont {Yu}\ \emph {et~al.}(2023)\citenamefont {Yu},
      \citenamefont {Xu}, \citenamefont {Qian}, \citenamefont {Qian},\ and\
      \citenamefont {Ji}}]{yu2023efficient}%
      \BibitemOpen
      \bibfield  {author} {\bibinfo {author} {\bibfnamefont {H.}~\bibnamefont
      {Yu}}, \bibinfo {author} {\bibfnamefont {Z.}~\bibnamefont {Xu}}, \bibinfo
      {author} {\bibfnamefont {X.}~\bibnamefont {Qian}}, \bibinfo {author}
      {\bibfnamefont {X.}~\bibnamefont {Qian}},\ and\ \bibinfo {author}
      {\bibfnamefont {S.}~\bibnamefont {Ji}},\ }\bibfield  {title} {\bibinfo
      {title} {Efficient and equivariant graph networks for predicting quantum
      {Hamiltonian}},\ }in\ \href {https://proceedings.mlr.press/v202/yu23i.html}
      {\emph {\bibinfo {booktitle} {International Conference on Machine
      Learning}}}\ (\bibinfo {organization} {PMLR},\ \bibinfo {year} {2023})\ pp.\
      \bibinfo {pages} {40412--40424}\BibitemShut {NoStop}%
    \bibitem [{\citenamefont {Fabrizio}\ \emph {et~al.}(2019)\citenamefont
      {Fabrizio}, \citenamefont {Grisafi}, \citenamefont {Meyer}, \citenamefont
      {Ceriotti},\ and\ \citenamefont {Corminboeuf}}]{fabrizio2019electron}%
      \BibitemOpen
      \bibfield  {author} {\bibinfo {author} {\bibfnamefont {A.}~\bibnamefont
      {Fabrizio}}, \bibinfo {author} {\bibfnamefont {A.}~\bibnamefont {Grisafi}},
      \bibinfo {author} {\bibfnamefont {B.}~\bibnamefont {Meyer}}, \bibinfo
      {author} {\bibfnamefont {M.}~\bibnamefont {Ceriotti}},\ and\ \bibinfo
      {author} {\bibfnamefont {C.}~\bibnamefont {Corminboeuf}},\ }\bibfield
      {title} {\bibinfo {title} {Electron density learning of non-covalent
      systems},\ }\href
      {https://pubs.rsc.org/en/content/articlehtml/2019/sc/c9sc02696g} {\bibfield
      {journal} {\bibinfo  {journal} {Chem. Sci.}\ }\textbf {\bibinfo {volume}
      {10}},\ \bibinfo {pages} {9424} (\bibinfo {year} {2019})}\BibitemShut
      {NoStop}%
    \bibitem [{\citenamefont {Qiao}\ \emph {et~al.}(2022)\citenamefont {Qiao},
      \citenamefont {Christensen}, \citenamefont {Welborn}, \citenamefont {Manby},
      \citenamefont {Anandkumar},\ and\ \citenamefont
      {Miller~III}}]{qiao2022informing}%
      \BibitemOpen
      \bibfield  {author} {\bibinfo {author} {\bibfnamefont {Z.}~\bibnamefont
      {Qiao}}, \bibinfo {author} {\bibfnamefont {A.~S.}\ \bibnamefont
      {Christensen}}, \bibinfo {author} {\bibfnamefont {M.}~\bibnamefont
      {Welborn}}, \bibinfo {author} {\bibfnamefont {F.~R.}\ \bibnamefont {Manby}},
      \bibinfo {author} {\bibfnamefont {A.}~\bibnamefont {Anandkumar}},\ and\
      \bibinfo {author} {\bibfnamefont {T.~F.}\ \bibnamefont {Miller~III}},\
      }\bibfield  {title} {\bibinfo {title} {Informing geometric deep learning with
      electronic interactions to accelerate quantum chemistry},\ }\href
      {https://www.pnas.org/doi/abs/10.1073/pnas.2205221119} {\bibfield  {journal}
      {\bibinfo  {journal} {Proc. Natl. Acad. Sci. U.S.A.}\ }\textbf {\bibinfo
      {volume} {119}},\ \bibinfo {pages} {e2205221119} (\bibinfo {year}
      {2022})}\BibitemShut {NoStop}%
    \bibitem [{\citenamefont {Shao}\ \emph {et~al.}(2023)\citenamefont {Shao},
      \citenamefont {Paetow}, \citenamefont {Tuckerman},\ and\ \citenamefont
      {Pavanello}}]{shao2023machine}%
      \BibitemOpen
      \bibfield  {author} {\bibinfo {author} {\bibfnamefont {X.}~\bibnamefont
      {Shao}}, \bibinfo {author} {\bibfnamefont {L.}~\bibnamefont {Paetow}},
      \bibinfo {author} {\bibfnamefont {M.~E.}\ \bibnamefont {Tuckerman}},\ and\
      \bibinfo {author} {\bibfnamefont {M.}~\bibnamefont {Pavanello}},\ }\bibfield
      {title} {\bibinfo {title} {Machine learning electronic structure methods
      based on the one-electron reduced density matrix},\ }\href
      {https://www.nature.com/articles/s41467-023-41953-9} {\bibfield  {journal}
      {\bibinfo  {journal} {Nat. Commun.}\ }\textbf {\bibinfo {volume} {14}},\
      \bibinfo {pages} {6281} (\bibinfo {year} {2023})}\BibitemShut {NoStop}%
    \bibitem [{\citenamefont {Kohn}(1996)}]{Kohn1996}%
      \BibitemOpen
      \bibfield  {author} {\bibinfo {author} {\bibfnamefont {W.}~\bibnamefont
      {Kohn}},\ }\bibfield  {title} {\bibinfo {title} {Density functional and
      density matrix method scaling linearly with the number of atoms},\ }\href
      {https://doi.org/10.1103/PhysRevLett.76.3168} {\bibfield  {journal} {\bibinfo
       {journal} {Phys. Rev. Lett.}\ }\textbf {\bibinfo {volume} {76}},\ \bibinfo
      {pages} {3168} (\bibinfo {year} {1996})}\BibitemShut {NoStop}%
    \bibitem [{\citenamefont {Prodan}\ and\ \citenamefont
      {Kohn}(2005)}]{Prodan2005}%
      \BibitemOpen
      \bibfield  {author} {\bibinfo {author} {\bibfnamefont {E.}~\bibnamefont
      {Prodan}}\ and\ \bibinfo {author} {\bibfnamefont {W.}~\bibnamefont {Kohn}},\
      }\bibfield  {title} {\bibinfo {title} {Nearsightedness of electronic
      matter},\ }\href {https://doi.org/10.1073/pnas.0505436102} {\bibfield
      {journal} {\bibinfo  {journal} {Proc. Natl. Acad. Sci. U.S.A.}\ }\textbf
      {\bibinfo {volume} {102}},\ \bibinfo {pages} {11635} (\bibinfo {year}
      {2005})}\BibitemShut {NoStop}%
    \bibitem [{\citenamefont {Soler}\ \emph {et~al.}(2002)\citenamefont {Soler},
      \citenamefont {Artacho}, \citenamefont {Gale}, \citenamefont {Garc\'ia},
      \citenamefont {Junquera}, \citenamefont {Ordej\'on},\ and\ \citenamefont
      {S\'anchez-Portal}}]{siesta2002}%
      \BibitemOpen
      \bibfield  {author} {\bibinfo {author} {\bibfnamefont {J.~M.}\ \bibnamefont
      {Soler}}, \bibinfo {author} {\bibfnamefont {E.}~\bibnamefont {Artacho}},
      \bibinfo {author} {\bibfnamefont {J.~D.}\ \bibnamefont {Gale}}, \bibinfo
      {author} {\bibfnamefont {A.}~\bibnamefont {Garc\'ia}}, \bibinfo {author}
      {\bibfnamefont {J.}~\bibnamefont {Junquera}}, \bibinfo {author}
      {\bibfnamefont {P.}~\bibnamefont {Ordej\'on}},\ and\ \bibinfo {author}
      {\bibfnamefont {D.}~\bibnamefont {S\'anchez-Portal}},\ }\bibfield  {title}
      {\bibinfo {title} {The {SIESTA} method for ab initio order-{N} materials
      simulation},\ }\href {https://doi.org/10.1088/0953-8984/14/11/302} {\bibfield
       {journal} {\bibinfo  {journal} {J. Phys.: Condens. Matter}\ }\textbf
      {\bibinfo {volume} {14}},\ \bibinfo {pages} {2745} (\bibinfo {year}
      {2002})}\BibitemShut {NoStop}%
    \bibitem [{\citenamefont {Ozaki}\ and\ \citenamefont
      {Kino}(2004)}]{openmx2004}%
      \BibitemOpen
      \bibfield  {author} {\bibinfo {author} {\bibfnamefont {T.}~\bibnamefont
      {Ozaki}}\ and\ \bibinfo {author} {\bibfnamefont {H.}~\bibnamefont {Kino}},\
      }\bibfield  {title} {\bibinfo {title} {Numerical atomic basis orbitals from
      {H} to {Kr}},\ }\href {https://doi.org/10.1103/PhysRevB.69.195113} {\bibfield
       {journal} {\bibinfo  {journal} {Phys. Rev. B}\ }\textbf {\bibinfo {volume}
      {69}},\ \bibinfo {pages} {195113} (\bibinfo {year} {2004})}\BibitemShut
      {NoStop}%
    \bibitem [{\citenamefont {Martin}(2004)}]{Martin2004}%
      \BibitemOpen
      \bibfield  {author} {\bibinfo {author} {\bibfnamefont {R.~M.}\ \bibnamefont
      {Martin}},\ }\href {https://doi.org/10.1017/CBO9780511805769} {\emph
      {\bibinfo {title} {Electronic Structure: Basic Theory and Practical
      Methods}}}\ (\bibinfo  {publisher} {Cambridge University Press, Cambridge,
      England},\ \bibinfo {year} {2004})\BibitemShut {NoStop}%
    \bibitem [{\citenamefont {Gilbert}(1975)}]{gilbert1975hohenberg}%
      \BibitemOpen
      \bibfield  {author} {\bibinfo {author} {\bibfnamefont {T.~L.}\ \bibnamefont
      {Gilbert}},\ }\bibfield  {title} {\bibinfo {title} {Hohenberg-{Kohn} theorem
      for nonlocal external potentials},\ }\href
      {https://journals.aps.org/prb/abstract/10.1103/PhysRevB.12.2111} {\bibfield
      {journal} {\bibinfo  {journal} {Phys. Rev. B}\ }\textbf {\bibinfo {volume}
      {12}},\ \bibinfo {pages} {2111} (\bibinfo {year} {1975})}\BibitemShut
      {NoStop}%
    \bibitem [{\citenamefont {Knizia}\ and\ \citenamefont
      {Chan}(2012)}]{knizia2012density}%
      \BibitemOpen
      \bibfield  {author} {\bibinfo {author} {\bibfnamefont {G.}~\bibnamefont
      {Knizia}}\ and\ \bibinfo {author} {\bibfnamefont {G.~K.-L.}\ \bibnamefont
      {Chan}},\ }\bibfield  {title} {\bibinfo {title} {Density matrix embedding:
      {A} simple alternative to dynamical mean-field theory},\ }\href
      {https://journals.aps.org/prl/abstract/10.1103/PhysRevLett.109.186404}
      {\bibfield  {journal} {\bibinfo  {journal} {Phys. Rev. Lett.}\ }\textbf
      {\bibinfo {volume} {109}},\ \bibinfo {pages} {186404} (\bibinfo {year}
      {2012})}\BibitemShut {NoStop}%
    \bibitem [{\citenamefont {Ozaki}\ and\ \citenamefont
      {Terakura}(2001)}]{ozaki2001convergent}%
      \BibitemOpen
      \bibfield  {author} {\bibinfo {author} {\bibfnamefont {T.}~\bibnamefont
      {Ozaki}}\ and\ \bibinfo {author} {\bibfnamefont {K.}~\bibnamefont
      {Terakura}},\ }\bibfield  {title} {\bibinfo {title} {Convergent recursive
      {O(N)} method for ab initio tight-binding calculations},\ }\href
      {https://journals.aps.org/prb/abstract/10.1103/PhysRevB.64.195126} {\bibfield
       {journal} {\bibinfo  {journal} {Phys. Rev. B}\ }\textbf {\bibinfo {volume}
      {64}},\ \bibinfo {pages} {195126} (\bibinfo {year} {2001})}\BibitemShut
      {NoStop}%
    \bibitem [{\citenamefont {Passaro}\ and\ \citenamefont
      {Zitnick}(2023)}]{escn2023}%
      \BibitemOpen
      \bibfield  {author} {\bibinfo {author} {\bibfnamefont {S.}~\bibnamefont
      {Passaro}}\ and\ \bibinfo {author} {\bibfnamefont {C.~L.}\ \bibnamefont
      {Zitnick}},\ }\bibfield  {title} {\bibinfo {title} {Reducing {SO}(3)
      convolutions to {SO}(2) for efficient equivariant {GNN}s},\ }in\ \href
      {https://proceedings.mlr.press/v202/passaro23a.html} {\emph {\bibinfo
      {booktitle} {International Conference on Machine Learning}}}\ (\bibinfo
      {publisher} {PMLR},\ \bibinfo {year} {2023})\ pp.\ \bibinfo {pages}
      {27420--27438}\BibitemShut {NoStop}%
    \bibitem [{\citenamefont {Liao}\ \emph {et~al.}(2024)\citenamefont {Liao},
      \citenamefont {Wood}, \citenamefont {Das},\ and\ \citenamefont
      {Smidt}}]{liao2024equiformerv2}%
      \BibitemOpen
      \bibfield  {author} {\bibinfo {author} {\bibfnamefont {Y.-L.}\ \bibnamefont
      {Liao}}, \bibinfo {author} {\bibfnamefont {B.~M.}\ \bibnamefont {Wood}},
      \bibinfo {author} {\bibfnamefont {A.}~\bibnamefont {Das}},\ and\ \bibinfo
      {author} {\bibfnamefont {T.}~\bibnamefont {Smidt}},\ }\bibfield  {title}
      {\bibinfo {title} {{EquiformerV2}: Improved equivariant transformer for
      scaling to higher-degree representations},\ }in\ \href
      {https://openreview.net/forum?id=mCOBKZmrzD} {\emph {\bibinfo {booktitle}
      {International Conference on Learning Representations}}}\ (\bibinfo {year}
      {2024})\BibitemShut {NoStop}%
    \bibitem [{\citenamefont {Ozaki}(2003)}]{openmx2003}%
      \BibitemOpen
      \bibfield  {author} {\bibinfo {author} {\bibfnamefont {T.}~\bibnamefont
      {Ozaki}},\ }\bibfield  {title} {\bibinfo {title} {Variationally optimized
      atomic orbitals for large-scale electronic structures},\ }\href
      {https://doi.org/10.1103/PhysRevB.67.155108} {\bibfield  {journal} {\bibinfo
      {journal} {Phys. Rev. B}\ }\textbf {\bibinfo {volume} {67}},\ \bibinfo
      {pages} {155108} (\bibinfo {year} {2003})}\BibitemShut {NoStop}%
    \bibitem [{\citenamefont {Hamann}\ \emph {et~al.}(1979)\citenamefont {Hamann},
      \citenamefont {Schl\"uter},\ and\ \citenamefont {Chiang}}]{ncpp1979}%
      \BibitemOpen
      \bibfield  {author} {\bibinfo {author} {\bibfnamefont {D.~R.}\ \bibnamefont
      {Hamann}}, \bibinfo {author} {\bibfnamefont {M.}~\bibnamefont {Schl\"uter}},\
      and\ \bibinfo {author} {\bibfnamefont {C.}~\bibnamefont {Chiang}},\
      }\bibfield  {title} {\bibinfo {title} {Norm-conserving pseudopotentials},\
      }\href {https://doi.org/10.1103/PhysRevLett.43.1494} {\bibfield  {journal}
      {\bibinfo  {journal} {Phys. Rev. Lett.}\ }\textbf {\bibinfo {volume} {43}},\
      \bibinfo {pages} {1494} (\bibinfo {year} {1979})}\BibitemShut {NoStop}%
    \bibitem [{\citenamefont {Perdew}\ \emph {et~al.}(1996)\citenamefont {Perdew},
      \citenamefont {Burke},\ and\ \citenamefont {Ernzerhof}}]{pbe1996}%
      \BibitemOpen
      \bibfield  {author} {\bibinfo {author} {\bibfnamefont {J.~P.}\ \bibnamefont
      {Perdew}}, \bibinfo {author} {\bibfnamefont {K.}~\bibnamefont {Burke}},\ and\
      \bibinfo {author} {\bibfnamefont {M.}~\bibnamefont {Ernzerhof}},\ }\bibfield
      {title} {\bibinfo {title} {Generalized gradient approximation made simple},\
      }\href {https://doi.org/10.1103/PhysRevLett.77.3865} {\bibfield  {journal}
      {\bibinfo  {journal} {Phys. Rev. Lett.}\ }\textbf {\bibinfo {volume} {77}},\
      \bibinfo {pages} {3865} (\bibinfo {year} {1996})}\BibitemShut {NoStop}%
    \end{thebibliography}
\end{document}